\documentstyle{article}
\newcommand{\dir}{.}
\newcommand{\figpage}[2]
{
     \LARGE
     \noindent
     \unitlength=1mm
     \begin{picture}(140,130)
     \put(0,0){
       \psfig{figure=\dir/#2,width=140mm,height=120mm}
     }
     \end{picture}
     \vfill
     \normalsize
     {\tt
       \noindent
       Figure #1 \\
       \figauth \\
     }
   \noindent
}
\newcommand{\xfigpage}[2]
{
     \LARGE
     \noindent
     \unitlength=1mm
     \begin{picture}(140,130)
     \put(0,0){
       \psfig{figure=\dir/#2,width=120mm,height=120mm}
     }
     \end{picture}
     \vfill
     \normalsize
     {\tt
       \noindent
       Figure #1 \\
       \figauth \\
     }
   \noindent
}
\begin{document}
\large
\baselineskip=0.8cm
\input psfig

\newcommand{\figauth}{F. Schmid, H. Lange, J. Chem. Phys.}
\title{\bf \LARGE 
Influence of the Head Group Size on the Direction of Tilt in Langmuir Monolayers
}
\author
{F. Schmid, H. Lange\\
Institut f\"ur Physik, Universit\"at Mainz, D55099 Mainz, Germany
}
\date{}
\maketitle

{\bf Abstract} - 

A model of rods with heads of variable size, which are confined
to a planar surface, is used to study the influence of the head group
size on tilted phases in Langmuir monolayers. Simple free energy 
considerations as well as exact zero temperature calculations
indicate that molecules with small head groups tilt towards next nearest 
neighbors, and molecules with larger head groups towards nearest neighbors.
This provides a possible explanation for recent experimental results, and
for details of the generic phase diagram for fatty acid monolayers.

\vfill

\newpage

\section{Introduction.}

Monolayers of simple amphiphiles at the air-water interface form a 
complex variety of condensed phases \cite{mono-p,kaganer}. A generic
phase diagram for fatty acids is shown in Figure 1.
The phase behavior appears to be driven mainly by the conformations of the
aliphatic tails of the molecules. However, head group interactions also play a 
role, especially in tilted phases. 
Shih {\em et al} \cite{shih2} and Fischer {\em et al} \cite{fischer} 
have recently performed systematic studies of mixtures of heneicosanic acid
and heneicosanol. They find that the $L_{2(h,d)}$ phases with tilt
towards nearest neighbors are gradually suppressed, as the alcohols are
added to the fatty acid monolayer, until they finally disappear at
alcohol concentrations above 35 \%. In pure heneicosanol monolayers,
only phases with tilt towards next nearest neighbors exist. 

According to a common picture, tilt order in monolayers is often induced by
a mismatch between the head group spacing and the chain diameter.
The size of the head groups plays a crucial role for this mechanism. 
From the observation that the zero pressure lattice constant is smaller in 
heneicosanol monolayers than in heneicosanic acid monolayers \cite{shih2},
one can deduce that the effective size of alcohol head groups in water
is smaller than that of fatty acid head groups. The results of Shih {\em et al}
and Fischer {\em et al} therefore suggest that the head group size also
determines the direction of tilt. Further investigations of surface pressure --
temperature phase diagrams for monolayers of fatty acids with esters
by Teer {\em et al} \cite{teer} are consistent with this picture.

We investigate this possibility in the framework of the simplest possible
theoretical model: Amphiphilic molecules are represented by rigid rods with
heads larger than the rod diameter. We discuss the model first using a
simple, general free energy {\em ansatz}. Our results are confirmed 
by exact zero temperature calculations for a system with Lennard-Jones 
interactions.

\section{Free Energy Considerations.}

We consider rigid rods of length $L$ with steric repulsive interactions
and longer range attractive interactions; the favorite distance between two
rods at given temperature is $r_t$. The rods have heads of radius $r_h$, 
which are confined to a planar surface and arranged on a distorted hexagonal 
lattice, with variable lattice constants $a$ and $b$ (Figure 2). 
Two cases are discussed, tilt in the direction of $a$ (nearest neighbors, NN) 
and tilt in the direction of $b$ (next nearest neighbors, NNN). 
In this geometry, rods have two different types of neighbors, 
labelled (i) and (ii) in the following.

The head size gives rise to the geometrical constraints
\begin{equation}
\label{cth}
\mbox{(i)} \qquad a \geq r_h \qquad 
\mbox{(ii)} \qquad (a^2+b^2)/4 \geq r_h^2.
\end{equation}
If the heads are small, $r_h \leq r_t$, the optimal configuration is one of
untilted rods arranged on an undistorted hexagonal lattice, with
the lattice constant $a=r_t$. The volume per rod of the rod layer is then 
given by $V_0 = \sqrt{3}/2 \: r_t^2 \: L$. 
If the heads are larger than the tails, the attractive interactions of 
the tails cause them to tilt. This may involve a distortion of the lattice; 
the surface per rod $A$ occupied by the monolayer increases, 
which is in turn penalized by the surface tension $\Sigma$ and the surface 
pressure $\Pi$.

Such considerations lead to the free energy {\em ansatz}
\begin{equation}
\label{free}
F = V \; \frac{1}{2 \kappa} (\frac{V_0}{V} -1)^2 + (\Sigma+\Pi) A 
\end{equation}
with the volume compressibility of the rods $\kappa$ and the volume per 
rod $V$.  The repulsive interactions between the rods cause additional
constraints
\begin{equation}
\label{ctn} 
\mbox{(i)} \quad  
{a \stackrel{>}{\sim} r_t \quad \atop a \cos \theta \stackrel{>}{\sim} r_t }
\qquad
\mbox{(ii)} \quad  
{(a^2+b^2 \cos^2 \theta)/4 \stackrel{>}{\sim} r_t^2 \atop
 (a^2 \cos^2 \theta + b^2)/4 \stackrel{>}{\sim} r_t^2}
\quad
{ \quad \mbox{(NNN)} \atop \quad \mbox{(NN)} }
\end{equation}
where $\theta$ is the tilt angle.
It is convenient to define the mismatch parameter $\delta = 1-r_h/r_t$, 
and to rewrite the free energy as
\begin{equation}
\label{lambda}
F \quad  \propto  \quad
\tilde{F} \equiv
\Lambda (\sqrt{\frac{V_0}{V}} - \sqrt{\frac{V}{V_0}})^2 + \frac{L A}{V_0}
\qquad \mbox{with} \qquad 
\Lambda = \frac{L}{2 \kappa (\Sigma + \Pi)}.
\end{equation}
Our task is to minimize $\tilde{F}$ with respect to the surface $A=ab/2$
and volume $V=A L \cos\theta > V_0$, under the constraints (\ref{cth}) and
(\ref{ctn}). The remaining independent model parameters are 
$\Lambda$ and $\delta$. Assuming that $k = \Lambda \delta$ is of order unity
or less, the minimum value of $\tilde{F}$ can be expanded in powers of 
$\delta$. Details of the calculation are presented in the appendix.

In layers with rods tilted towards next nearest neighbors, the volume
per rod cannot be optimized, because the distance to the neighbors 
(i) is too large, $a > r_h > r_t$. The free energy is minimized at
$A=\sqrt{3} r_h^2$ and takes the value
\begin{equation}
\label{fnnn}
\tilde{F}_{NNN} = 1+ (2+\frac{4}{9} k) \delta + (1-\frac{28}{27} k) \delta^2
+ \cdots
\end{equation}
The head lattice is not distorted. This is an effect of
the ``hard'' constraints of eqn (\ref{cth}), {\em i.e.} the hard
core interactions between heads. If the heads interact {\em via} a softer
potential, the lattice gets slightly distorted in the direction of the tilt.

In layers with rods tilted towards nearest neighbors, the volume per rod
can be optimized, but at the cost of an increased
surface energy. Surface and volume contributions have to be balanced. 
For large $k$, $k \gg 1$, the volume term determines most of
the behavior, and one gets
\begin{equation}
\label{fnn1}
\tilde{F}_{NN} \stackrel{k\gg1}{=}
\left\{ \begin{array}{lr}
1+4 \delta - 6 \delta^2 + \cdots & 
\quad \mbox{:} \quad \delta > \delta_0 \quad \mbox{(a)} \\
1+(4-1/k) \delta - (6-16/k+10/k^2) \delta^2 + \cdots & 
\quad \mbox{:} \quad \delta < \delta_0 \quad \mbox{(b)}
\end{array} \right.
\end{equation}
with $\delta_0 = \sqrt{3/2}-1 = 0.225$. In this case, the head lattice is 
distorted, and the volume per rod is minimal ($V=V_0$) in the limit
$k \to \infty$ or $\delta > \delta_0$.
For small $k$ on the other hand, $k \ll 1$, the surface term dominates. 
The heads form an undistorted lattice, and the free energy is given by
\begin{equation}
\label{fnn2}
\tilde{F}_{NN} \stackrel{k \ll 1}{=}  1+(2+k) \delta + (1-k) \delta^2 + \cdots
\end{equation}
In the intermediate regime, $k \sim 1$, the smaller of the two solutions 
(\ref{fnn1}) or (\ref{fnn2}) applies. Note that the free energy (\ref{fnn2})
is always larger than (\ref{fnnn}). Hence rods tilt towards next
nearest neighbors at $k \ll 1$, or generally if distortions
of the head lattice are not allowed.

Suppose now that $\Lambda$ is kept fixed, and the head size is gradually 
increased. Rods with small heads, $\delta < \delta_{c1} = 0$, are untilted. 
If the heads are just slightly larger than the tails, $0<\delta < 1/\Lambda$, 
the case $k<1$ applies, and the rods tilt towards their next nearest 
neighbors. On increasing the head size (and if $\Lambda$ is sufficiently
large), a swiveling transition to a phase with tilt towards nearest neighbors 
takes place. The transition point $\delta_{c2}$ is located by equating 
eqn (\ref{fnn1}) with eqn (\ref{fnnn}). 
One thus gets a sequence of phase transitions at 
\begin{equation}
\label{dc}
\delta_{c1} = 0 \qquad \mbox{and} \qquad 
\delta_{c2} = k_c/\Lambda \quad \mbox{with} \quad
k_c = 3.93 
\end{equation}
to lowest order of $\delta$ ($\Lambda >  k_c/\delta_0)$.
Note that the transition at $\delta_{c1}$ from the untilted to the tilted 
state, does not depend on $\Lambda$. This is again a consequence of the 
hard core interactions between the heads. If the heads are soft, they are
squeezed together at high pressures ({\em i.e.}, small $\Lambda$), and the 
rods stand up. Note also that the free energy (\ref{free}) is temperature
dependent {\em via} the temperature dependence of the compressibility 
$\kappa$ and the surface tension $\Sigma$. However, the model cannot
be expected to reproduce thermal properties of a monolayer,
since the hard core interactions between the heads are basically
athermal.

\section{Exact Ground State Calculations.}

In order to test these predictions, we have studied numerically the ground
state of a system of rods with Lennard-Jones interactions between tails 
and repulsive soft core interactions between heads. The interaction 
energy of two rods at grafting distance $\vec{r}$, 
with tilt direction $\vec{e}_L$, is given by
\begin{equation}
E(\vec{r},\vec{e}_L) = \int_0^L \!\int_0^L dl \: dl' \: V_{LJ}(d) + V_{h}(r),
\end{equation}
where $d$ is the distance between elements $dl$ and $dl'$ on the rods,
\mbox{$d = | \vec{r} + (l-l') \vec{e}_L |$}, and the potentials are taken
to be
\begin{eqnarray}
V_{LJ}(d) &=& \left\{
\begin{array}{lcl}
d^{-12} - 2 d^{-6} + 0.031 &:& d \le 2 \\
0 &:& d > 2
\end{array}
\right. \\
V_{h}(r) &=& \left\{
\begin{array}{lcl}
(\sigma/r)^{12} - 2 (\sigma/r)^{6} +1 &:& r \le \sigma \\
0 &:& r > \sigma
\end{array}
\right. 
\end{eqnarray}
The potentials $V_{LJ}$ and $V_h$ are cut at $d=2$ and $r=\sigma$, 
respectively, in order to make the problem more tractable.
The total energy $E_t$ is the sum over all nonzero pair interactions, 
{\em i.e.}, the interactions of a rod with up to 24 pairs \cite{balashov}. 
It is minimized with fixed molecular area $A$ by numerical methods. 
The surface pressure is obtained using $\Pi = - dE_{t,min}/dA$.

The lattice constant and the compressibility of a system of infinitely
long rods can be calculated analytically. We identify these
parameters with the rod diameter $r_t$ and the volume compressibility 
$\kappa$ ($r_t=0.9333$ and $\kappa = 0,0184$), and $\sigma$ with the head 
size $r_h$. This allows a quantitative comparison of the numerical results 
with the predictions of the previous section.

The phase diagram for chain length $L=5$ is shown in Figure 3. With increasing 
head size, one finds the sequence of phases predicted in section 1: 
Untilted (U), Tilt towards next nearest neighbors (NNN) and towards nearest 
neighbors (NN). Due to the soft core interactions between the
heads, the first transition U-NNN becomes pressure dependent. Apart from this
detail, the results are in qualitative and reasonable quantitative
agreement with the predictions of eqn (\ref{dc}).

\section{Discussion.}

In sum, we find that tilt order can be induced in systems of rods 
with grafted heads by a mismatch of head size and rod diameter.
Rods with large heads tilt towards nearest neighbors,
and rods with heads just slightly larger than the rod diameter
tilt towards next nearest neighbors. 
The critical head size, at which the swiveling transition occurs, increases 
with pressure and decreases with the rod length. If distortions of the 
hexagonal head lattice are not allowed, rods always tilt towards their next 
nearest neighbors. 

This is in agreement with previous theoretical predictions. Scheringer 
et al \cite{scheringer} report that rods grafted on an undistorted hexagonal 
lattice tilt towards next nearest neighbors. 
Kaganer et al \cite{kaganer2} find that this remains true, even if one allows 
for lattice distortions, in systems of pure rods (no heads) which are
forced to tilt by the constraint of a fixed homogeneous surface density
(see also \cite{frank}).
Some authors study the effect that tilt towards nearest neighbors can be 
induced by additional, attractive interactions between the rods 
and the surface \cite{kaganer2,cai}. The present work indicates that it might
simply be a consequence of the larger head group size.

Our effect offers a simple explanation for the observations of Shih et al 
\cite{shih2} and Fischer et al \cite{fischer}, discussed in the introduction. 
It also provides an interpretation for earlier experimental findings. 
In particular, the fact that phases with tilt towards next nearest neighbors
appear at high surface pressures (see Figure 1), comes out quite naturally.

Shih et al \cite{shih1} have studied monolayers of heneicosanic acid in the 
presence of calcium ions in the subphase, and systematically varied the pH of 
the subphase. At low pH, they find a sequence of phases with tilt towards 
nearest neighbors, tilt towards next nearest neighbors, and no tilt, as a 
function of pressure. On increasing the pH, the phase tilted towards next 
nearest neighbors moves down to lower pressures, and finally replaces the 
phase with tilt towards nearest neighbors. At even higher pH, the monolayer 
is untilted at all pressures. Assuming that the effective size of the 
COOH${}^-$ head groups is reduced at high pH -- {\em e.g.}, due to more 
efficient screening of the electrostatic interactions at higher concentration 
of positive ions in the subphase -- this result corresponds exactly to the
predictions of our model.

To conclude, our simple considerations provide an explanation for a remarkable 
variety of experimental observations. Obviously, the real interactions between 
head groups in Langmuir monolayers are much more complicated than those
assumed here. Nevertheless, the success of our model suggests that the 
influence of head groups on the microscopic structure of Langmuir monolayers is
to a large extent simply determined by their size.

\section*{Acknowledgement.}

We thank K. Binder, C. Stadler, F.M. Haas and R. Hilfer for helpful discussions.

\section*{Appendix}

We wish to minimize the free energy (\ref{free}) with the constraints
(\ref{cth}) and (\ref{ctn}). For tilt towards next nearest
neighbors, the solution can be written down immediately:
$a = r_h$, $b = \sqrt{3} r_h$,
$\cos \theta = \sqrt{(4-(1+\delta))^2)/3 (1+\delta)^2}$,
and 
\begin{equation}
\tilde{F}_{NNN} =
(1+\delta)^2 + \Lambda ( 
\frac{3 + 4 (1+\delta)^2 - (1+\delta)^4}
     {\sqrt{3} (1+\delta) \sqrt{4 - (1+\delta)^2}}     -2 )
\end{equation}
The expansion of $\tilde{F}_{NNN}$ in powers of $\delta$ yields (\ref{fnnn}).

For tilt towards nearest neighbors, different solutions are possible.
If the volume term dominates, one achieves $V=V_0$ with
$\cos \theta = r_t/a$, $b = \sqrt{3} r_t$ 
and minimizes the surface $A$ under this constraint with
$a/r_t = \sqrt{4 (1+\delta)^2-3}$.
\begin{equation}
\label{fnnv}
\tilde{F}_{NN}^{(V)} = \sqrt{4 (1+\delta)^2-3}.
\end{equation}
In this case, the head lattice is distorted.
Expansion of $\tilde{F}_{NN}^{(V)}$ gives (\ref{fnn1}.a). 

If the surface term dominates, the head lattice is not distorted,
($b = \sqrt{3} r_h$, $a = r_h$) and with this constraint 
the volume is minimal for
$\cos \theta = r_t/a$.
\begin{equation}
\label{fnna}
\tilde{F}_{NN}^{(A)} = (1+\delta)^2 + \Lambda
\frac{\delta^2}{(1+\delta)}
\end{equation}
Expansion yields (\ref{fnn2}). 

If none of the two terms dominates,
the solution lies between (\ref{fnnv}) and (\ref{fnna}): 
$b = \sqrt{3} r_t y$ and 
$ a/r_t = \sqrt{4 (1+\delta)^2 - 3 y^2}$ with $1<y<1+\delta$
($\cos \theta = r_t/a$). 
\begin{equation} 
\tilde{F}_{NN} = \min_{1<y<1+\delta} {\cal F}(y) \quad \mbox{with} \quad
{\cal F}(y) = y \sqrt{4 (1+\delta)^2 - 3 y^2 } + \Lambda \frac{(y-1)^2}{y} 
\end{equation}
Note that $d{\cal F}/dy|_{y=1} {>\atop <} 0$ for 
$ \delta {>\atop <} \delta_0 = \sqrt{3/2}-1$. If $\delta > \delta_0$,
$\delta_0$, the function ${\cal F}(y)$ has no minimum in the interval
$]1,1+\delta[$, and $\tilde{F}_{NN}$ is either given by 
${\cal F}(1)=\tilde{F}_{NN}^{(A)}$, or 
${\cal F}(1+\delta)=\tilde{F}_{NN}^{(V)}$.
If $\delta<\delta_0$, ${\cal F}(y)$ has a minimum. 
The surface term still dominates at small $\Lambda$ 
($\Lambda \stackrel{<}{\sim} 1/\delta$), 
$\tilde{F}_{NN}=\tilde{F}_{NN}^{(A)}$. 
At larger $\Lambda$, no closed expression for $\tilde{F}_{NN}$ can be given,
but an expansion in powers of $\delta$ is possible:
The minimum of ${\cal F}(y)$ is found at $y = 1+\alpha \delta$ with
\begin{equation}
\alpha = \frac{1}{k} + \frac{21-16k}{2k^2}\: \delta 
+ \frac{4(40-42k+9k^2)}{k^3}\: \delta^2 + \cdots
\end{equation}
($k=\Lambda \delta$), which leads to eqn (\ref{fnn1}.b).

\newpage

\section*{Figure Captions}

\begin{description}

\item[Figure 1:]
Part of the generic phase diagram for fatty acid monolayers. 
Solid lines are first order transitions, broken lines are second order.
The amphiphile molecules are tilted towards next nearest neighbors
in the phases $Ov$ and $L_2'$, towards nearest neighbors
in $L_{2d}$ and $L_{2h}$, and untilted in $S$ and $LS$. 
In the $S$, $L_2'$ and $L_{2h}$ phases, the backbones of the chains 
are ordered, the other phases are rotator phases. 
Increasing the length of the aliphatic chains shifts the phase 
diagram to higher temperatures (after \cite{kaganer}).

\item[Figure 2:]
Side view of the rigid rod model (top) 
and top view of the head lattice (bottom). We consider tilt directions
in the direction of $a$ (nearest neighbors), or $b$ (next nearest
neighbors. (i) and (ii) mark the different types of direct neighbors 
in this geometry.

\item[Figure 3:]
Ground state phase diagram of the model of section 3 (rods with
Lennard-Jones interactions) in the plane of head size $\sigma$ vs 
pressure $\Pi$, at rod length $L=5$. Phases are: tilt towards next nearest 
neighbors (NNN), towards nearest neighbors (NN), untilted (U). 
Dashed lines indicate the prediction of eqn (\ref{dc}) with $\Sigma = 6.7$.

\end{description}

\newpage
\pagestyle{empty}

\xfigpage{1}{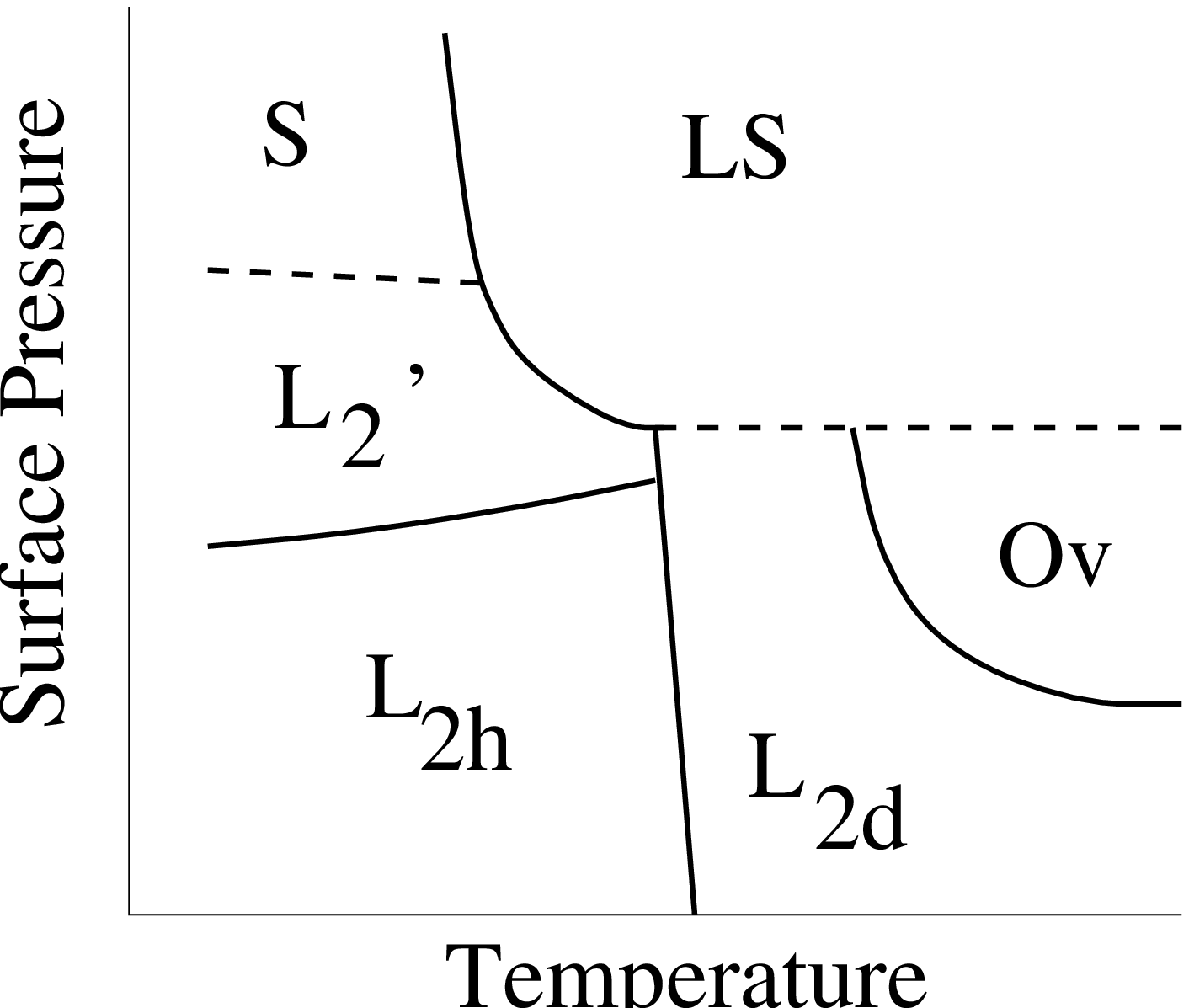}

\xfigpage{2}{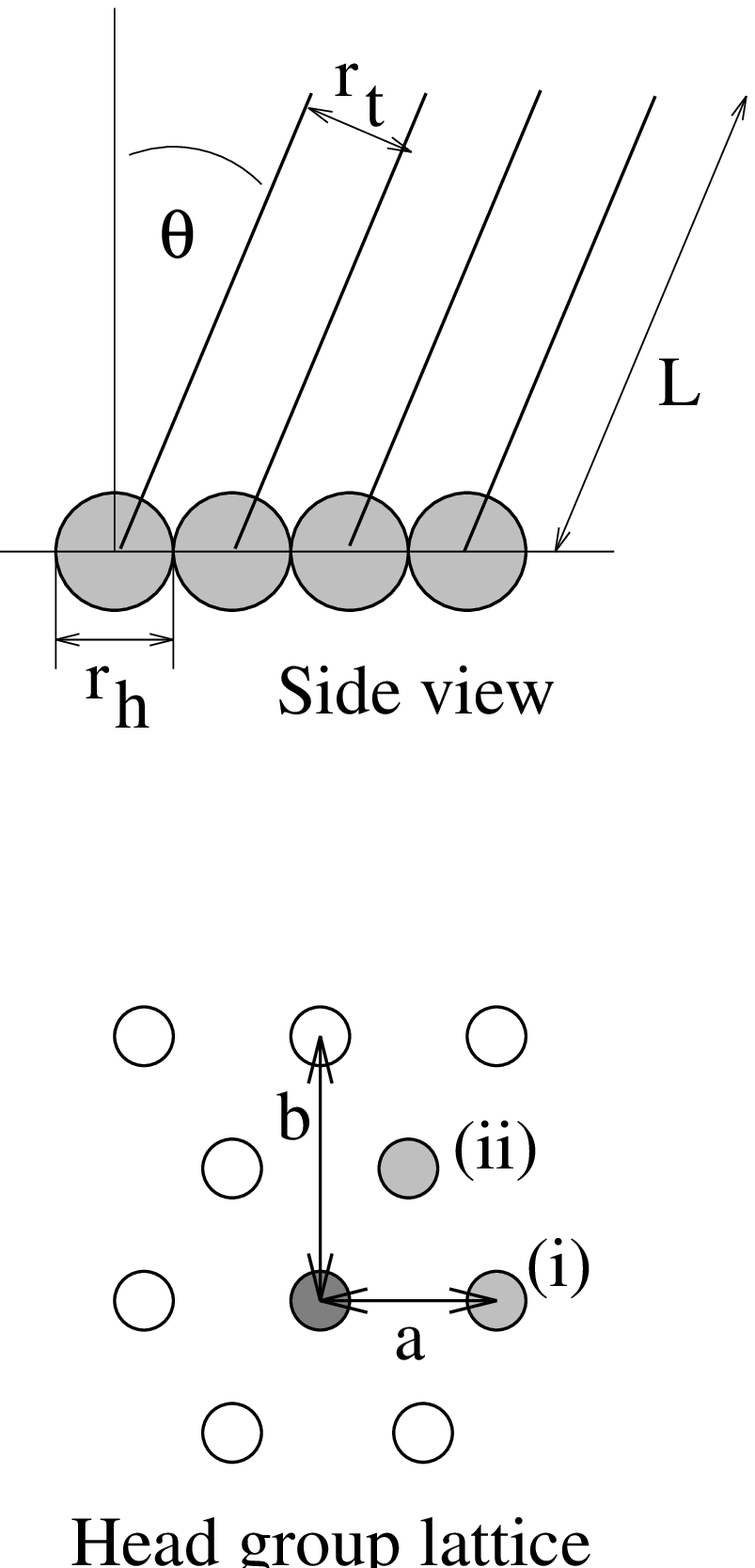}

\figpage{3}{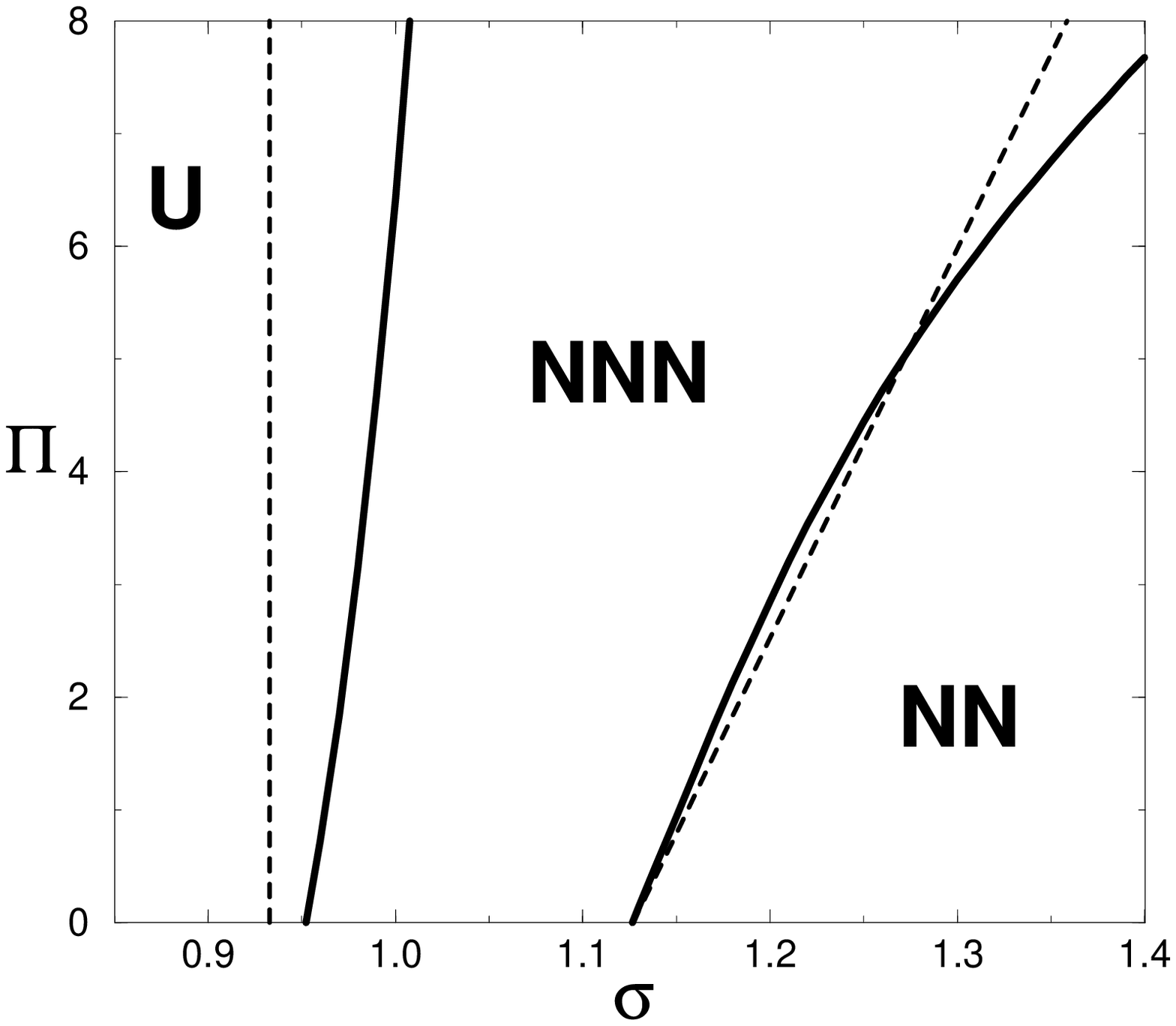}

\end{document}